\documentclass[a4paper,11pt]{article}
\usepackage{pos}
\usepackage{caption}
\usepackage{subcaption}
\usepackage[mathlines]{lineno}
\title{Nuclear modification of $\pi^0$ production\\in pO and OO collisions with ALICE}
\ShortTitle{Nuclear modification of $\pi^0$ production in pO and OO collisions with ALICE}

\author[a]{Nicolas Strangmann}

\onbehalf{for the ALICE Collaboration}

\affiliation[a]{Institut für Kernphysik, Johann Wolfgang Goethe-Universität Frankfurt, Frankfurt, Germany}

\emailAdd{nicolas.strangmann@cern.ch}

\abstract{
Energy loss of high-momentum partons is one of the key signatures of the quark--gluon plasma formed in heavy-ion collisions.
Through measurements of the nuclear modification factor, $R_{\rm AA}$, where a strong suppression of high-$p_{\rm T}$ particle yields is observed, parton energy loss has been confirmed in Pb--Pb collisions at the LHC.
However, in small collision systems, like pp and p--Pb, energy loss has not yet been confirmed.

In this contribution, a measurement of the $\pi^0$ $R_{\rm AA}$ in OO and pO collisions at $\sqrt{s_{\rm NN}} = 5.36$ TeV and $\sqrt{s_{\rm NN}} = 9.62$~TeV, respectively, is presented. These results are based on the measured $\pi^0$ $p_{\rm T}$ spectra in OO, pO, and pp collisions using the ALICE electromagnetic calorimeter.
The measurement of the $R_{\rm pO}$ allows for precise constraints on cold-nuclear matter effects. In combination with the $R_{\rm OO}$, this allows for a separation of cold and hot nuclear matter contributions and enables a quantitative determination of partonic energy loss in OO collisions.
The measurements are compared to theoretical model calculations that include both cold and hot nuclear matter effects.\\

The results presented in this contribution are based on a letter accepted by PRL~\cite{OOELossPaper}.
}

\FullConference{13th International Conference on Hard and Electromagnetic Probes of High-Energy Nuclear Collisions (HardProbes2026)\\
21-26 June 2026\\
Nashville, Tennessee, USA\\}

\begin{document}
\maketitle

\section{Physics context}

The observation of collective behavior in pp and p--Pb collisions at the LHC challenges the role of these small systems as clean references for measurements in larger collision systems and has extended the search for quark--gluon plasma (QGP) signatures beyond heavy-ion collisions~\cite{JFSmallSystemReview}.
Parton energy loss ("jet quenching") provides a key test of how far these signatures extend towards smaller systems: it has been firmly established in Au--Au and Pb--Pb collisions, while no unambiguous evidence has been found in pp or p--Pb.
A key observable for studying parton energy loss is the nuclear modification factor, which compares particle production in nuclear collisions with a scaled pp reference.
For production cross sections $\mathrm{d}\sigma/\mathrm{d}p_{\mathrm{T}}$ of a given particle in minimum bias collisions, it is defined as
\begin{equation}
\label{eq:RAA}
R_{\mathrm{AA}} = \frac{1}{A^{2}}
\frac{\left.\mathrm{d}\sigma_{\mathrm{AA}}/\mathrm{d}p_{\mathrm{T}}\right.}
     {\left.\mathrm{d}\sigma_{\mathrm{pp}}/\mathrm{d}p_{\mathrm{T}}\right.},
\quad
R_{\mathrm{pA}} = \frac{1}{A}
\frac{\left.\mathrm{d}\sigma_{\mathrm{pA}}/\mathrm{d}p_{\mathrm{T}}\right.}
     {\left.\mathrm{d}\sigma_{\mathrm{pp}}/\mathrm{d}p_{\mathrm{T}}\right.},
\end{equation}
where $A$ is the nuclear mass number, and $A^{2}$ and $A$ count the number of individual nucleon--nucleon pairs of the two colliding nuclei.
Measurements of high-$p_{\mathrm{T}}$ hadron production in Au--Au and Pb--Pb collisions show strong suppression relative to this scaled reference, whereas corresponding p--A measurements show no suppression.
This contrast leaves the onset of observable parton energy loss at intermediate system sizes unresolved.

The first OO collisions delivered by the LHC in July 2025 provide an opportunity to explore this intermediate system size, thereby constraining the onset of parton energy loss. 
However, interpreting $R_{\mathrm{OO}}$ requires control over cold nuclear matter (CNM) effects, which modify particle production even without a hot medium and are commonly described through nuclear parton distribution functions (nPDFs).
These modifications are poorly constrained for oxygen, leading to substantial uncertainties in extracting hot-medium effects from the data.
A complementary measurement of $R_{\mathrm{pO}}$ therefore provides an experimental constraint on CNM effects and helps to distinguish their contribution from possible hot-medium effects in OO collisions.
The double ratio $R_{\mathrm{OO}}/R_{\mathrm{pO}}^{2}$ exploits this constraint to largely cancel CNM contributions and their associated nPDF uncertainties, increasing sensitivity to parton energy loss~\cite{Jonas:2026yoz}.
The squared denominator accounts approximately for nuclear modifications in both incoming nuclei in OO collisions, compared with a single nucleus in pO.

\section{Experimental setup and data sets}

The measurement uses minimum-bias OO and pO collisions recorded with ALICE in July 2025 at $\sqrt{s_{\mathrm{NN}}}=5.36$ and $9.62$~TeV, respectively.
Events are triggered by coincident signals in the FT0-A and FT0-C detectors~\cite{FITRef}, with the primary vertex required to lie within 10 cm of the nominal interaction point along the beam axis.
Neutral pions are reconstructed through their $\pi^{0}\rightarrow\gamma\gamma$ decay using the electromagnetic calorimeter (EMCal)~\cite{EMCalPerformance}, which covers $|\eta|<0.7$ and an azimuthal interval of $173^{\circ}$.

The analyzed integrated luminosities are $(1.87\pm0.04)$ nb$^{-1}$ for OO and $(3.6\pm0.1)$ nb$^{-1}$ for pO.
The pp reference data for OO, taken at the same center-of-mass energy per nucleon of $\sqrt{s}=5.36$~TeV, correspond to $(22.8\pm0.6)$ nb$^{-1}$. An additional sample at $13.6$~TeV, corresponding to $(137\pm4)$ nb$^{-1}$, enables an interpolation of the measured spectra to the pO collision energy.

\section{Neutral-pion reconstruction and production cross sections}

Photon candidates are reconstructed by clustering adjacent EMCal cells, each with a deposited energy of at least $100$~MeV.
Clusters are required to have an energy above 600 MeV, and shower-shape selections reduce hadronic contamination.
A timing requirement of $|t_{\mathrm{clus}}|<15$ ns suppresses out-of-bunch pileup.
Identical cluster selections are applied across all collision systems and energies.

Neutral pions are reconstructed from the invariant-mass distributions of photon-candidate pairs within each event.
The combinatorial background is described by a weighted combination of mixed-event and cluster-rotation templates, accounting for uncorrelated and correlated contributions.
Their weights are determined through a template fit, and the raw $\pi^{0}$ yield is extracted by integrating around the signal peak after background subtraction.
The yields are corrected for detector acceptance, reconstruction efficiency, secondary contributions, and the $\pi^{0}\rightarrow\gamma\gamma$ branching ratio, and normalized to the integrated luminosity.
Acceptance and efficiency corrections are obtained from PYTHIA~8 simulations~\cite{PYTHIA} with a GEANT4-based detector response~\cite{GEANT}.
The luminosity determination uses minimum-bias trigger cross sections measured in van der Meer scans and accounts for pileup from multiple collisions per triggered bunch crossing.

The pp reference at $\sqrt{s}=9.62$~TeV is obtained using a power-law energy dependence fitted in each $p_{\mathrm{T}}$ interval to previous $\pi^{0}$ cross-section measurements at 2.76, 7, 8, and 13~TeV~\cite{Pi0276,Pi07,Pi08,Pi013}.
The spectra at 5.36 and 13.6~TeV are scaled to the target energy and averaged.
A correction estimated with PYTHIA~8 matches the reference to the pO rapidity interval, changing it by less than 1\%.

\begin{figure}[tbp]
\centering
\includegraphics[width=0.49\textwidth]{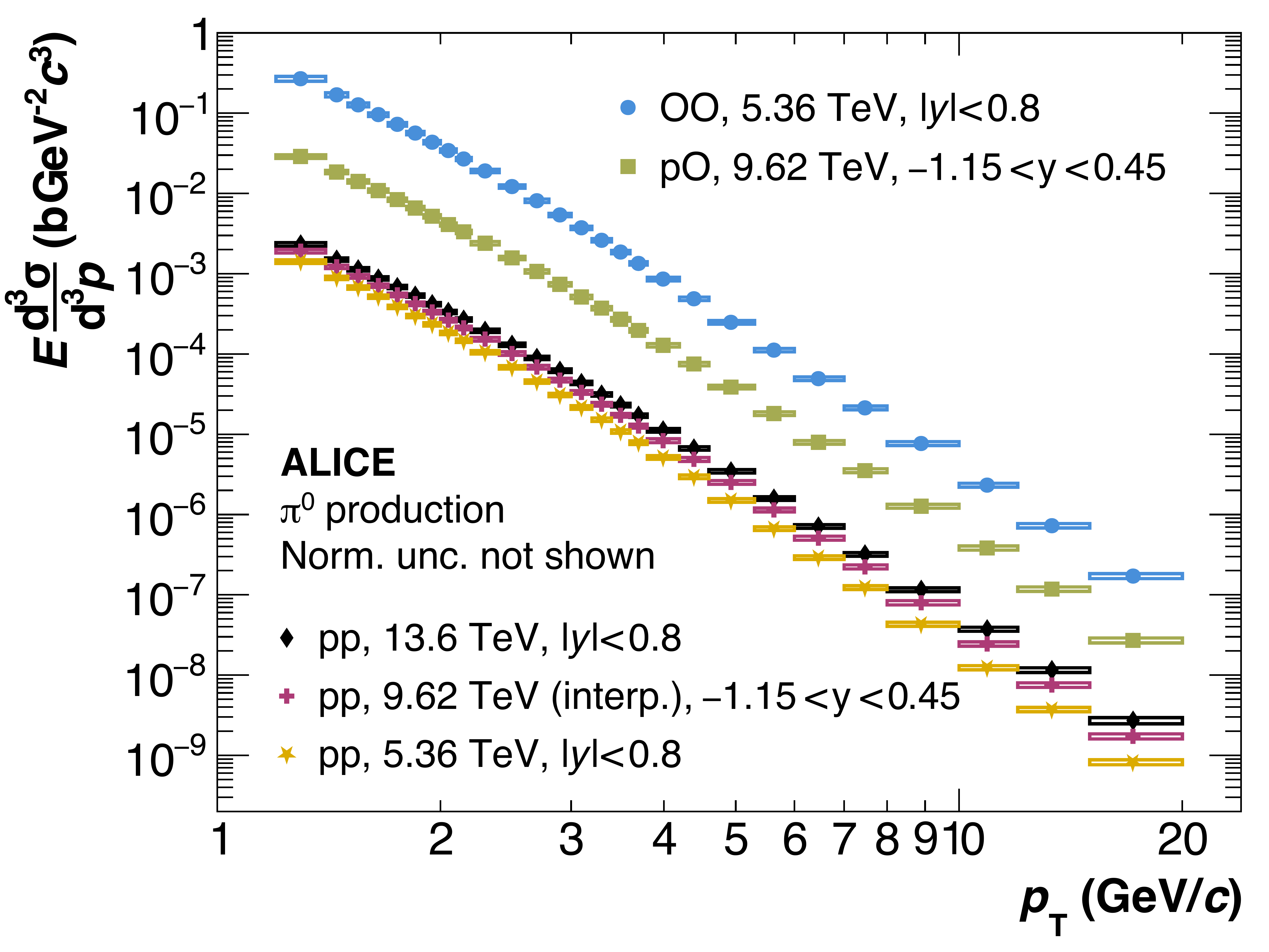}
\caption{Invariant $\pi^{0}$ production cross sections in OO collisions at $\sqrt{s_{\mathrm{NN}}}=5.36$ TeV, pO collisions at $\sqrt{s_{\mathrm{NN}}}=9.62$ TeV, and pp collisions at $\sqrt{s}=5.36$ and $13.6$ TeV, together with the interpolated pp reference at $\sqrt{s}=9.62$ TeV.
Vertical bars and boxes represent statistical and systematic uncertainties, respectively.}
\label{XSec}
\end{figure}

Figure~\ref{XSec} shows the measured invariant $\pi^{0}$ production cross sections in OO, pO, and pp collisions, together with the interpolated pp reference at $\sqrt{s}=9.62$ TeV.
The spectra cover $1.2<p_{\mathrm{T}}<20$ GeV/$c$ within $|y|<0.8$ for OO and the measured pp samples, and $-1.15<y<0.45$ for pO and its interpolated reference.

Systematic uncertainties are evaluated by varying the cluster selections and signal-extraction parameters.
These variations are applied simultaneously to all contributing spectra to account for correlations in $R_{\mathrm{OO}}$, $R_{\mathrm{pO}}$, and $R_{\mathrm{OO}}/R_{\mathrm{pO}}^{2}$.
Many detector-related uncertainties consequently cancel, leaving energy calibration and, for observables involving pO, reference interpolation as leading contributions.
The relative systematic uncertainties are approximately 5\% for the cross sections, 3\% for the nuclear modification factors and 5\% for the double ratio, with luminosity normalization uncertainties of 2.0\% for OO, 2.7\% for pO, and 2.6\% for each pp sample treated separately.

\section{Nuclear modification in pO and OO collisions}

\begin{figure}
     \centering
     \begin{subfigure}[b]{0.49\textwidth}
         \centering
         \includegraphics[width=\textwidth]{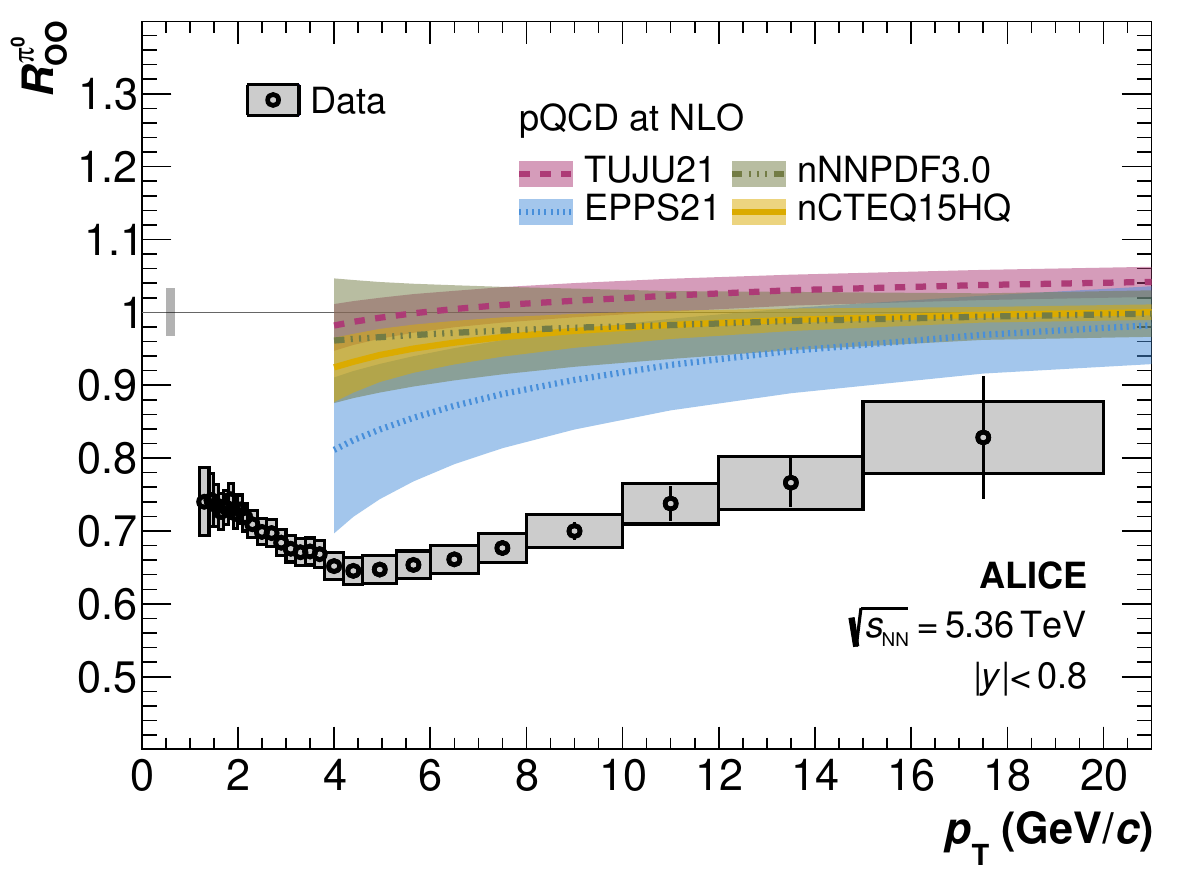}
         \caption{Comparison of the measured $R_{\rm OO}$ to nPDF predictions of CNM effects without energy loss~\cite{Jonas:2026yoz}.}
         \label{fig:ROO_nPDF}
     \end{subfigure}
     \hfill
     \begin{subfigure}[b]{0.49\textwidth}
         \centering
         \includegraphics[width=\textwidth]{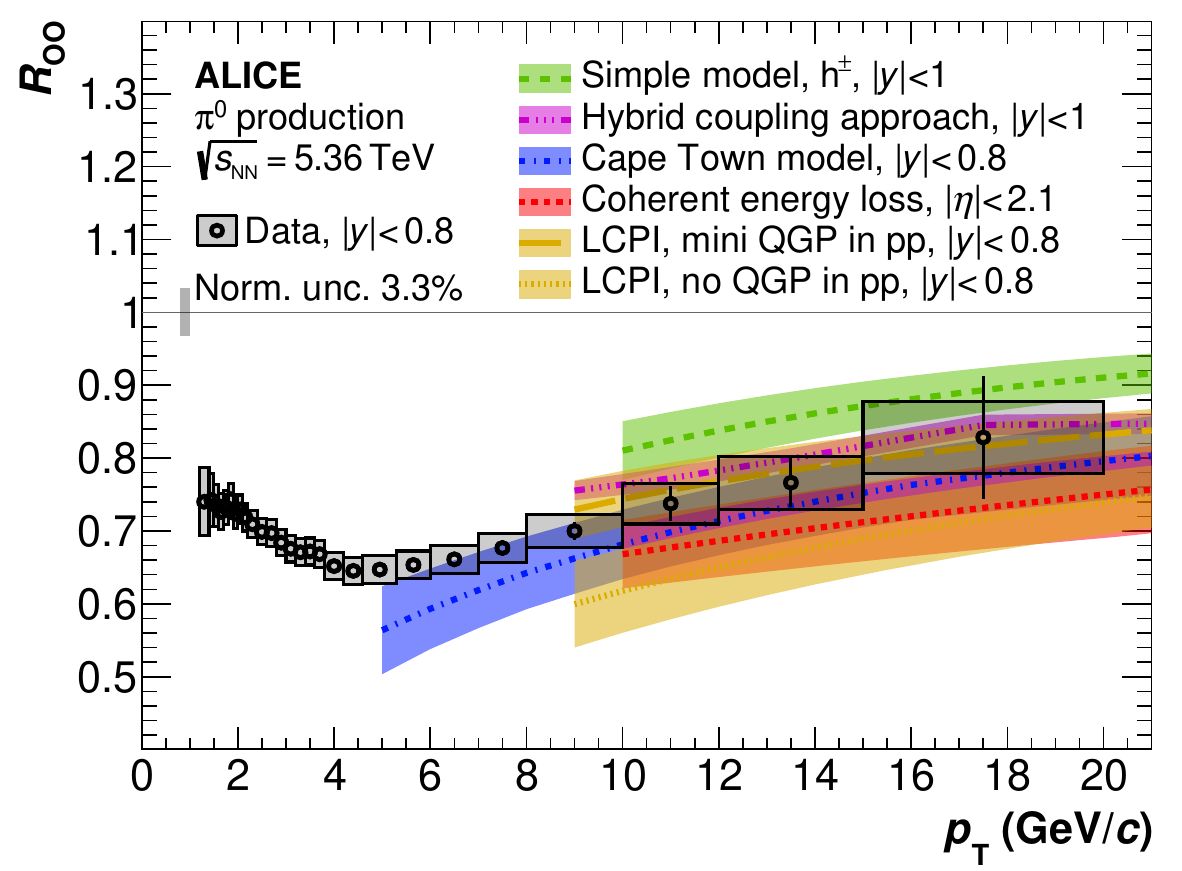}
         \caption{Comparison of the measured $R_{\rm OO}$ to energy-loss model predictions~\cite{ELossTheory1, ELossTheory2, ELossTheory3, ELossTheory4}.}
         \label{fig:ROO_ELossTheory}
     \end{subfigure}
        \caption{Measured nuclear modification factor $R_{\rm OO}$ compared to nPDF predictions without energy loss (left) and energy-loss models (right).
        Vertical bars and boxes represent statistical and systematic uncertainties, respectively.
Boxes around unity indicate luminosity normalization uncertainties.}
        \label{fig:ROO_Theory}
\end{figure}

Figure~\ref{fig:ROO_Theory} shows the measured neutral-pion nuclear modification factor, $R_{\rm OO}$, compared with calculations with and without parton energy loss.
The data exhibit a clear suppression relative to unity with a significance of 9.2$\sigma$.
However, suppression can also arise from CNM effects, as illustrated by the next-to-leading-order pQCD calculations using different nPDF sets~\cite{Jonas:2026yoz} in Fig.~\ref{fig:ROO_nPDF}.
While the central predictions from TUJU21, nNNPDF3.0, and nCTEQ15HQ remain within approximately 5\% of unity, EPPS21 predicts suppression of up to 20\%.
The large associated nPDF uncertainties limit the significance of suppression beyond CNM expectations to $1.8\sigma$ when EPPS21 is used as a baseline.
Significances are evaluated by generating pseudo-data under the no-energy-loss hypothesis using the experimental and theoretical covariance matrices, and determining how often these exhibit a larger covariance-weighted suppression than observed in the data.

The energy-loss models shown in Fig.~\ref{fig:ROO_ELossTheory} all describe the data within uncertainties, despite implementing parton energy loss through very different mechanisms~\cite{ELossTheory1, ELossTheory2, ELossTheory3, ELossTheory4}.
However, these calculations also use different nPDF baselines, and the spread of their predictions is comparable to that of the CNM baselines in Fig.~\ref{fig:ROO_nPDF}.
Differences between the predictions therefore cannot be attributed unambiguously to the energy loss mechanisms, preventing quantitative discrimination between them using $R_{\rm OO}$ alone.
This motivates the use of the measured $R_{\rm pO}$ to constrain CNM effects and the double ratio $R_{\rm OO}/R_{\rm pO}^{2}$ to reduce their contribution and associated uncertainties.

\begin{figure}
     \centering
     \begin{subfigure}[b]{0.49\textwidth}
         \centering
         \includegraphics[width=\textwidth]{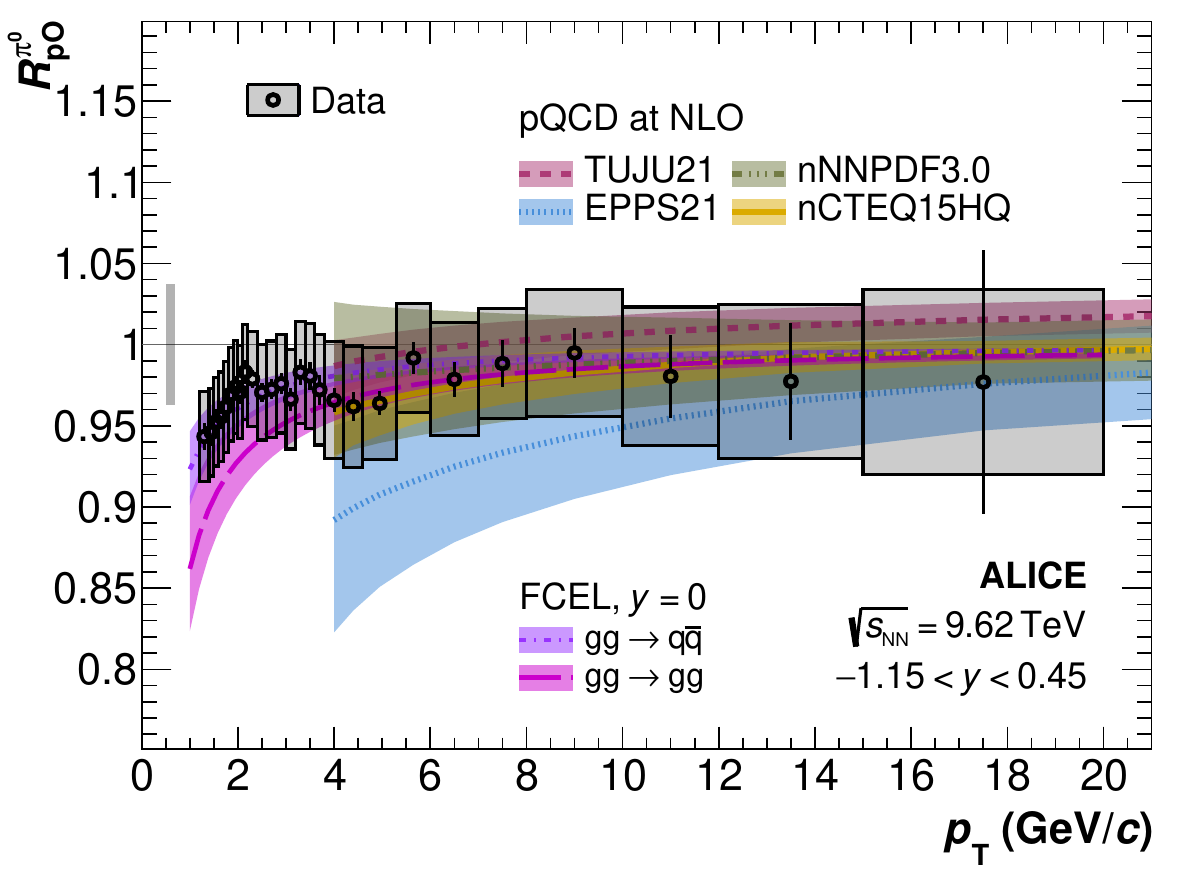}
         \caption{$R_{\rm pO}$ compared with nPDF-based CNM predictions~\cite{Jonas:2026yoz} and the fully coherent energy loss model calculations~\cite{FCEL}.}
         \label{fig:RpO}
     \end{subfigure}
     \hfill
     \begin{subfigure}[b]{0.49\textwidth}
         \centering
         \includegraphics[width=\textwidth]{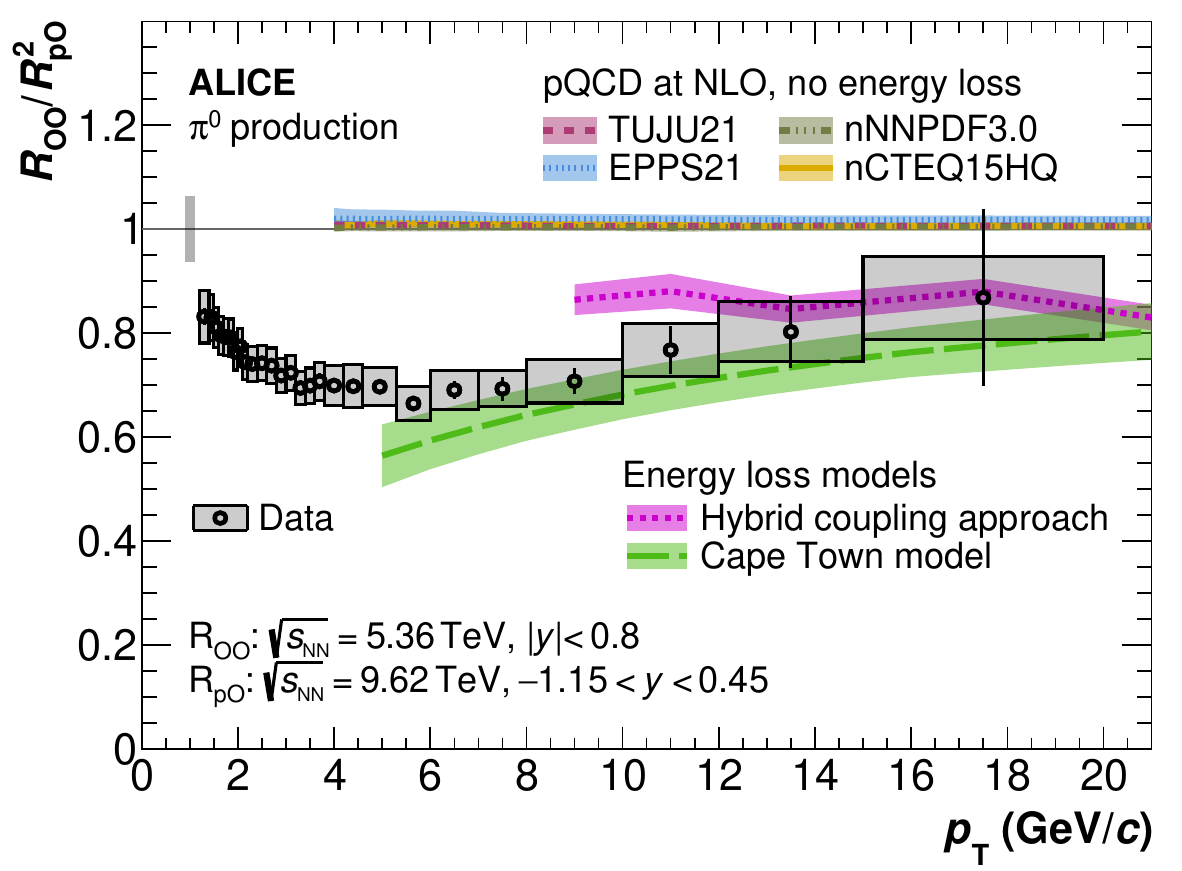}
         \caption{$R_{\rm OO}/R_{\rm pO}^{2}$ compared with predictions without energy loss~\cite{Jonas:2026yoz} and two energy-loss models~\cite{ELossTheory1, ELossTheory3}.}
         \label{fig:ROORpO2}
     \end{subfigure}
        \caption{Neutral-pion nuclear modification factor $R_{\rm pO}$ (left) and double ratio $R_{\rm OO}/R_{\rm pO}^{2}$ (right).
Vertical bars and boxes represent statistical and systematic uncertainties, respectively.
Boxes around unity indicate luminosity normalization uncertainties.}
\end{figure}

Figure~\ref{fig:RpO} shows the measured $R_{\rm pO}$, which is compatible with unity over the full $p_{\mathrm{T}}$ range, indicating no significant suppression.
The data are described within uncertainties by the nPDF-based CNM calculations~\cite{Jonas:2026yoz}, although their large uncertainties limit discrimination between the different nPDF sets.
Calculations by the fully coherent energy loss model~\cite{FCEL}, which describe CNM effects through coherent gluon radiation from scattered partons, are also compatible with the measurement.
This first measurement of $R_{\rm pO}$ provides an experimental constraint on oxygen nPDFs and a data-driven CNM baseline for OO collisions.

Combining the measured $R_{\rm OO}$ and $R_{\rm pO}$ into the double ratio $R_{\rm OO}/R_{\rm pO}^{2}$ largely cancels CNM contributions and their associated nPDF uncertainties~\cite{Jonas:2026yoz}.
In pO collisions, the higher collision energy probes smaller momentum fractions $x$, while the shifted rapidity probes larger $x$, so the two effects partially compensate.
The resulting nPDF-based predictions are compatible with unity, with uncertainties on the order of 1\%, illustrating the reduced sensitivity to nPDF modifications.
In contrast, the measured double ratio shown in Fig.~\ref{fig:ROORpO2} exhibits significant suppression.
The deviation from the nNNPDF3.0 baseline corresponds to $4.9\sigma$, with slightly larger significances obtained for the other nPDF sets, providing strong evidence for parton energy loss in OO collisions.
The reduced sensitivity to CNM effects enables quantitative comparisons with two energy-loss models~\cite{ELossTheory1,ELossTheory3}, both of which describe the magnitude and $p_{\mathrm{T}}$ dependence of the suppression within uncertainties.

\section{Summary}

Neutral-pion production cross sections are measured in OO collisions at $\sqrt{s_{\rm NN}}=5.36$~TeV, pO collisions at $\sqrt{s_{\rm NN}}=9.62$~TeV, and pp collisions at $\sqrt{s}=5.36$ and $13.6$~TeV.
While $R_{\rm OO}$ exhibits significant suppression and $R_{\rm pO}$ is compatible with unity, the double ratio $R_{\rm OO}/R_{\rm pO}^{2}$ largely cancels CNM contributions and reveals suppression beyond expectations without energy loss at a significance of $4.9\sigma$, providing evidence for parton energy loss in OO collisions.
All calculations without energy loss fail to describe this suppression, whereas two models including energy loss reproduce it within uncertainties.\\

This contribution was supported by BMFTR and the Helmholtz Association.

\end{document}